\documentclass[aps,prl,reprint,groupedaddress]{revtex4-1}

\usepackage{graphicx}

\begin{document}


\title{Multimode propagation in phononic crystals\\with overlapping Bragg and hybridization effects}


\author{C. Cro\"{e}nne}
\affiliation{Department of Physics and Astronomy, University of Manitoba, Winnipeg, Manitoba R3T 2N2, Canada}

\author{E. J. S. Lee}
\affiliation{Department of Physics and Astronomy, University of Manitoba, Winnipeg, Manitoba R3T 2N2, Canada}

\author{J. H. Page}
\email[]{john.page@umanitoba.ca}
\affiliation{Department of Physics and Astronomy, University of Manitoba, Winnipeg, Manitoba R3T 2N2, Canada}


\date{\today}

\begin{abstract}
Unusual dispersion properties are observed in a phononic crystal of nylon rods in water when the lattice constant is adjusted so that Bragg and hybridization gaps overlap in frequency. On the basis of experimental and numerical analyses of time-dependent transmission and spatial field maps, the presence of two coexisting propagation modes of similar amplitude is demonstrated near the resonance frequency. This phenomenon is attributed to the coupling of the rod resonances arranged in a triangular lattice, with phase shifts driven by the Bragg condition.
\end{abstract}


\maketitle

Mesoscopic phononic structures present remarkable possibilities for manipulating the dispersion of coherent wave transport through the structures, thereby dramatically altering the frequency dependence of the velocity and attenuation. Consequently, during the past several years, there have been a growing number of papers devoted to the study of such structures, especially phononic crystals containing periodic arrangements of inclusions embedded inside various matrix materials \cite{DeyAMPC13}. One of the most basic goals is to generate band gaps in the dispersion of the material in order to inhibit wave transport in certain frequency ranges, using mechanisms such as Bragg scattering \cite{KusPRB94} or the hybridization between a resonant mode of the scatterers and a propagating mode of the matrix \cite{PsaPRB02}.
However, even though both mechanisms lead to the formation of low transmission frequency bands in finite thickness materials, the resulting behavior is still quite different, with hybridization gaps often exhibiting negative values of the group velocity, whereas inside Bragg gaps this parameter is positive (and larger than in the matrix material) \cite{YangPRL02,CroAIP11}. Another distinguishing feature is that Bragg gaps require periodicity, while hybridization gaps persist in disordered structures \cite{StiPRL08,CowPRB11}. Additionally, it has been shown that, with strong material parameter contrast between the inclusions and the matrix, or with carefully designed resonant scatterers, hybridization gaps can be driven low enough in frequency to enable the representation of the structures as homogeneous effective media with unique (and often remarkable) acoustic or elastic metamaterial properties \cite{LiuSci00,LiPRE04,YanPRL08}.

In this Letter, we explore different unusual behavior in which hybridization gaps also play a key role.  By driving the frequency of the hybridization gap inside the first Bragg gap of a phononic crystal, we demonstrate and explain a novel type of atypical dispersion effect due to the coupling between these two mechanisms. In previous studies, the emphasis has been on showing that combined hybridization and Bragg gaps can have exceptional properties in terms of band gap width and/or depth \cite{PagZK05,LerAPL09,BilAPL13}, but no evidence of strikingly atypical dispersion characteristics was found.  Here we show that by tuning the resonant frequency, the transmitted phase above resonance can be shifted by multiples of $2\pi$, leading to switching between apparently different dispersion curves, and that the transmission coefficient can differ by orders of magnitude in its minimum value at the band-gap center.  We explain this unusual behavior by demonstrating the co-existence at the same frequency of two propagating modes with different wavevectors, with the competition between these modes driving the switching behavior that is observed near the resonant frequency. The existence of these two modes is established through analysis of both experiments and simulations, giving insight into the character of each mode and the influence of both resonance and lattice symmetries.  The possibility of tuning the competition between these modes dynamically offers a new approach for controlling both phase and amplitude of the transmitted signals through phononic crystals.

The phononic crystals (PnC) investigated were made from 0.46-mm-diameter nylon rods, readily available from commercial fishing line, arranged in a 2D triangular lattice and surrounded by water. The rods were positioned 
inside a support structure with top and bottom plates in which holes were drilled to ensure accurate positioning of the rods. Crystal interfaces were oriented perpendicular to the ${\Gamma}M$ direction, and only normal incidence was considered. The crystal thickness ranged from one to fourteen unit cells. Nylon density, longitudinal and shear velocities are respectively 1150 kg/m$^3$, 2500 m/s and about 1000 m/s, with this last parameter depending quite significantly on experimental conditions. 
Wave propagation through these crystals was measured using pulsed ultrasonic experiments, which were conducted in a temperature-controlled water tank, using pairs of identical transducers (Panametrics) spanning the frequency range from 0.3 to 2.2 MHz. By taking a Fourier transform of the measured time signals, the complex transmission spectra (both amplitude and phase) were obtained, yielding the dispersion curves and the frequency dependence of the 
transmission coefficient. 

\begin{figure}
\includegraphics[width=8.5cm]{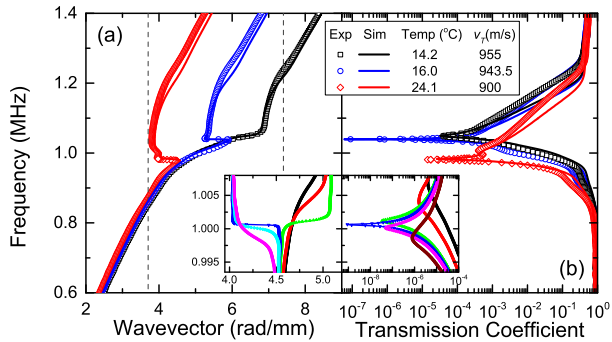}%
\caption{\label{Atypical}(color online) Atypical dispersion curves (a) and intensity transmission coefficients (b) measured (symbols) and simulated (lines) from the phase and amplitude of the transmission through a 5-cell-thick phononic crystal of nylon rods in water. Results for three temperatures are shown, with experimental data during swelling of the nylon rods 
in the insert. (In the insert, the transmission axis ranges from $5\times10^{-9}$ to $1\times10^{-4}$, and a 1 m/s change in $v_T$ is sufficient to switch between the left- and right-bending dispersion curves at the crossover point.)
The dashed vertical lines indicate the first and second Brillouin zone boundaries.}
\end{figure}

This crystal has the interesting property of exhibiting both hybridization and Bragg gaps in the same frequency range. For small lattice periods, the first Bragg gap appears at frequencies higher than the first resonant mode of the nylon rods, resulting into two distinct stop bands near 1.0 and 1.5 MHz that exhibit the characteristic group velocity signatures of hybridization and Bragg gaps (negative and large positive velocities, respectively). When the lattice constant $a$ is increased to 0.98 mm, the first Bragg gap is brought to lower frequencies and overlaps with the hybridization gap. The overlap is further controlled  by exploiting the strong dependence of the nylon shear velocity $v_T$ on experimental conditions, which allow the frequency of the hybridization gap to be finely tuned by changing the temperature, by swelling the nylon via water absorption through prolonged immersion in the water tank, or by varying the tension.  Under these conditions, we find that this crystal then exhibits remarkable transmission properties, both in phase and magnitude, with the interaction between hybridization and Bragg effects introducing phase rotations which change the effective wavenumber seen in the higher frequency dispersion branch.  An example of this atypical dispersion is shown in Fig.~\ref{Atypical} for a 5-layer-thick crystal at three representative temperatures. Anomalous switching behavior is also found when the number of layers in the slab is varied \cite{CroAIP11}.

\begin{figure}
\includegraphics[width=8.5cm]{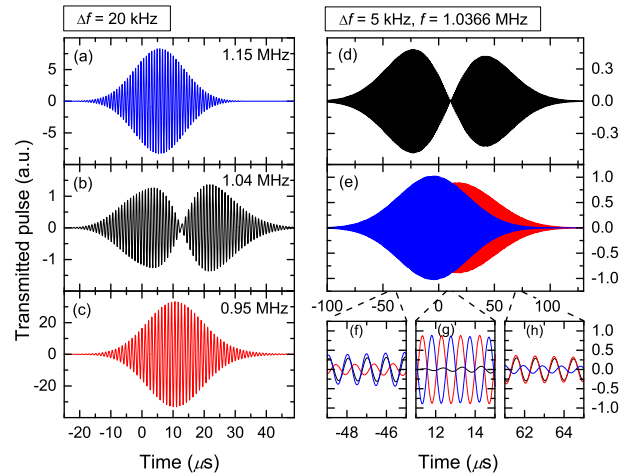}
\caption{\label{TimeExp}(color online) Transmitted experimental time signals through a 6-cell thick slab. (a,b,c) Filtered time signals 
with filters of same width ${\Delta}f$ = 20 kHz, but different central frequencies. (d, e) Time signals obtained by the modal decomposition procedure (original filtered signal in black, individual modes in blue and red). (f,g,h) Zoomed views of (d, e) at selected times.}
\end{figure}

\begin{figure}
\includegraphics[width=85mm]{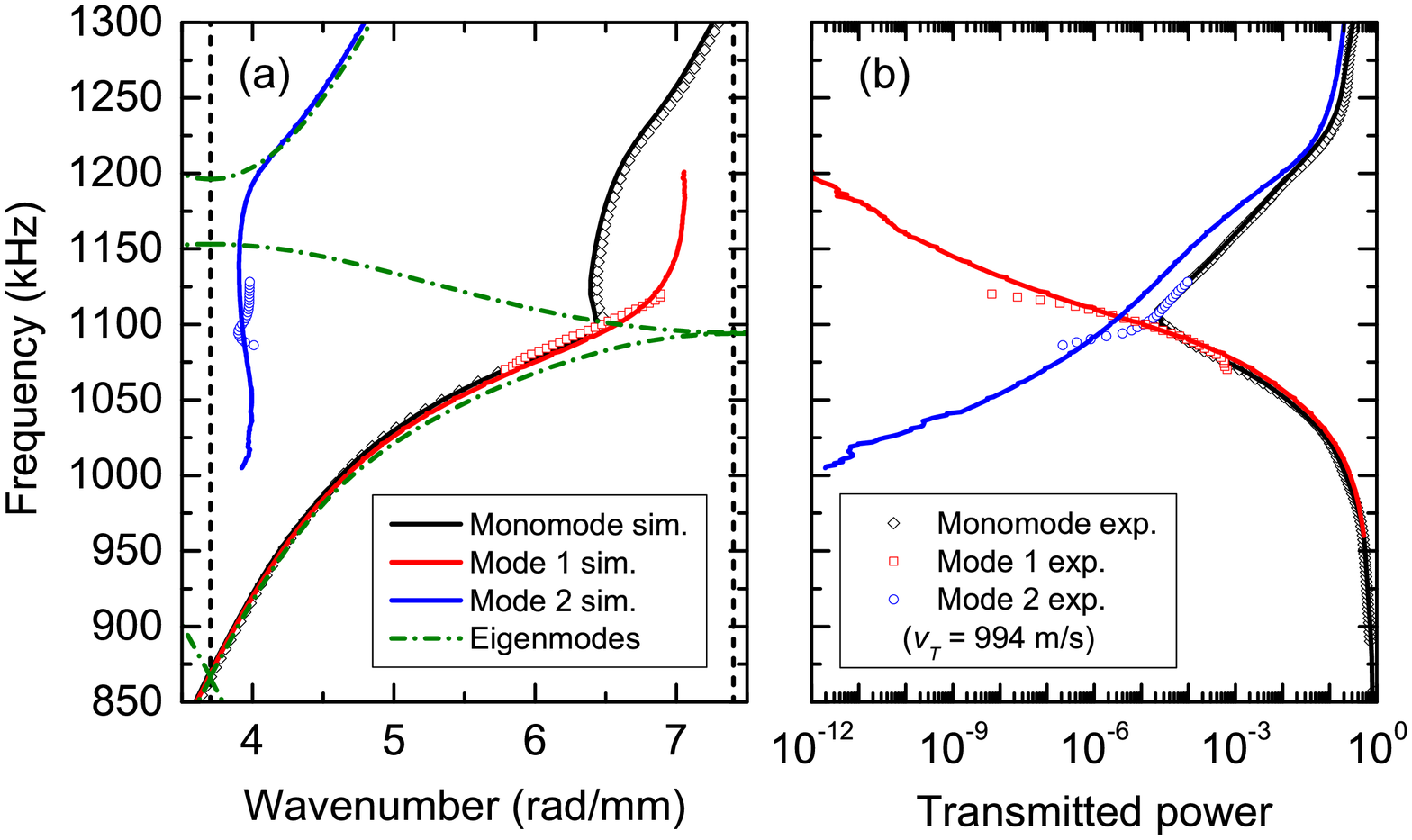}%
\caption{\label{DiagDisp}(color online) Experimental (symbols) and simulated (solid lines) frequency dependence of (a) the wavenumber and (b) the power transmission coefficient for the two individual modes (red and blue), and for the case when a single mode is assumed (black), with a 6-cell-thick slab. Vertical dashed lines correspond to the edges of the first two Brillouin zones. Panel (a) also shows the band diagram obtained when considering an infinite crystal (dot-dashed green lines).}
\end{figure}

The origin of this anomalous behavior is revealed by digitally filtering the transmitted experimental time signals with different narrow-bandwidth Gaussian frequency filters.  Figure \ref{TimeExp}(a, b, c) shows that the transmitted pulses are Gaussian over most of the frequency range, except around the resonance frequency. Near resonance, the observed signal shape [Fig. \ref{TimeExp}(b)] reveals a signature of multimode propagation, with two interfering pulses created by two propagation modes with different phase velocities, attenuations and group delays. With a fitting procedure, these individual modes can be retrieved. For instance, the signal shown in Fig. \ref{TimeExp}(d) can be separated into the two signals of Fig. \ref{TimeExp}(e). The overall shape of the transmitted pulse is then easily explained by inspecting the modes at different times (Fig. \ref{TimeExp}(f)): in the middle of the pulse, the signal becomes very low because the two modes are almost exactly out of phase. By repeating the fitting procedure with different frequency filters, dispersion properties of the individual modes can be extracted, as shown in Fig. \ref{DiagDisp} (symbols). The range of frequencies where this procedure can be applied is limited not only by the attenuation of the modes, but also by their phase difference, with the fitting method working most reliably when the two modes are out of phase with each other.  

To further investigate this multimode propagation effect, numerical simulations were performed using the ATILA finite element code \cite{ATILA}. A 2D domain was considered, with a plane wave of chosen frequency normally incident on a PnC slab of finite thickness, with periodic boundary conditions on the sides. Nylon losses were taken into account by adding an imaginary part to its bulk modulus, equal to 5.5 \% of the real part. To avoid including near field effects in the scattering parameters of the slab, transmitted and reflected fields were measured 4 mm away from the slab. Using the simulated complex transmission spectrum, attenuation and wavenumber can be calculated directly for a wide range of frequencies. However, since the harmonic simulation results can be affected by the internal reflections at the boundaries of the slab, a conventional inversion method \cite{FokPRB07,CroIEEE10} was preferred to retrieve the complex wavenumber, exploiting both reflection and transmission spectra.

For the modal decomposition, simulated displacement and pressure fields obtained for a 30-cell-thick slab were first interpolated on a square grid with a width of $a$ and a depth of $\sqrt{3}a/2$ centered on each rod in the slab, which allows building data sets with field values at the same position relative to the rod in each cell. Every data set was then fitted to a superposition of two arbitrary plane waves, for each frequency. Several optimization steps were then added to converge on the average complex wavevector values, as well as field maps, for both modes. With this procedure, two modes with imaginary wavevectors up to 2.2 mm$^{-1}$ could be fitted, without making any assumption about the nature of the modes.

The numerical modal decomposition result (Fig.~\ref{DiagDisp}) is in good agreement with the experimental one in the frequency range where both can be obtained. It shows that the two modes can be treated as extensions of the bands located below and above the gap. Indeed, the band structure of the infinite crystal, calculated with COMSOL Multiphysics, confirms the frequencies of the two passbands around the gap (dot-dashed green lines in Fig.~\ref{DiagDisp}). Since the eigenvalues are obtained in the real wavevector, complex frequency domain, this simulation does not give access to solutions in the frequency region where the two complex wavevector modes overlap. It should also be noted that this type of simulation reveals an additional resonance-related mode inside the gap, from 1100 to 1150 kHz. However, inspection of the corresponding field maps shows that this mode is deaf, i.e. anti-symmetric with respect to the propagation direction, and thus not excitable with normally incident plane waves.

Contrary to the usual hybridization effect in acoustic systems where losses are inevitable, the two passbands do not connect inside the gap to form a single dispersion branch with negative group velocity, but instead create a band where two modes with different wavevectors coexist.  This two-mode behavior is in sharp contrast to the case of a randomly disordered sample of the same nylon rods at the same rod concentration, where only one mode was found experimentally near 1 MHz \cite{LeeUSE09}, pointing to the important role of the interaction between Bragg and resonance-related effects in governing the competition between the two modes in our crystals. It is also interesting to compare our observations with experiments and theory for random dispersions of plastic spheres in a fluid, where an additional slowly propagating mode due to interfacial waves was found over a wide frequency range \cite{CowPRB11}.  While this 3D system also exhibits two-mode propagation, the character of the two modes is very different in our 2D crystal case.  Here the faster mode is strongly modified by the presence of the Brillouin zone boundary, forming a dispersion branch close to the one seen for Bragg gaps, whereas the slower mode crosses the first Brillouin zone boundary (around 870 kHz) without forming a gap. Additionally, the relative amplitude of the two modes has also been strongly modified, with a slow mode dominating the transmitted signal in a much larger frequency range, thus enhancing the multimode behavior.

The transmitted power spectra shown in Fig. \ref{DiagDisp}(b) are calculated using the average pressure fields obtained from the modal decomposition. They confirm that transmission in the lower and upper parts of the frequency spectrum is due respectively to the first and the second retrieved modes. The frequency where the transmission is equal for the two modes is very close to the frequency where a strong dependence of the effective monomode wavenumber on the slab thickness can be observed. This confirms the idea that the unusual behavior in transmission is due to the presence of two modes with similar amplitude but very different wavenumbers.

\begin{figure}
\includegraphics[width=85mm]{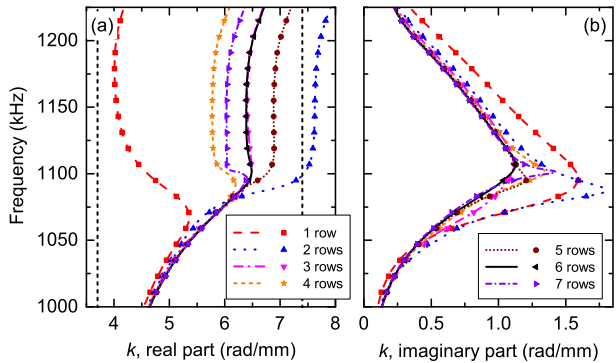}%
\caption{\label{DiagDisp1to7}(color online) Frequency dependence of (a) the real and (b) the imaginary parts of the wavenumber $k$ extracted from the simulations of 1 to 7-cell- thick slabs (lines) and reconstructed from the modal decomposition results (symbols). Vertical dashed lines show the edges of the first two Brillouin zones.}
\end{figure}

Indeed, from the modal decomposition results, the apparent monomode wavevectors for different slab thicknesses can be recalculated, provided that we have a good approximation of the transmission coefficients at the input interface, to represent the coupling between the incoming plane wave and the individual modes. If we base this approximation on the value of the average pressure field in the first cell for each mode (normalized by the input field), good qualitative agreement with the numerical monomode results can be obtained, with similar phase shifts at the resonance frequency and the same number of phase rotations. Additionally, if we use the coefficients at the input interface as fitting parameters, excellent quantitative agreement can be achieved for all thicknesses, as shown in Fig. \ref{DiagDisp1to7}. This shows that the coupling mechanism for the two individual modes is independent of the slab thickness.

\begin{figure}
\includegraphics[width=85mm]{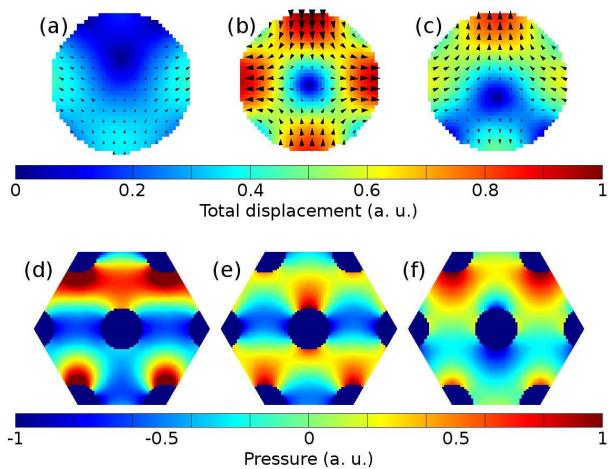}%
\caption{\label{Fields}(color online) Displacement (a, b, c) and pressure (d, e, f) amplitude field maps at 1105 kHz in the third cell of a slab: (a, d) total field, (b, e) first mode and (c, f) second mode (propagation direction is along the vertical axis). Displacement magnitude is shown both in the colormap and the arrow size. Pressure and displacement units are independent
.}
\end{figure}

Insight into the character of each mode can be gained from their displacement and pressure field maps, shown at 1105 kHz in Fig.~\ref{Fields}. At that frequency, their attenuation is equal. Note that the modes do not exhibit a large difference in terms of energy localization within the cell, so that neither mode can be identified as an interfacial mode.  In fact, both modes involve deformations of the rods from a circular to oval cross section (i.e., a quadrupolar resonance), even though the first mode has a more symmetric shape. The two modes also differ by the phase of the quadrupolar deformation in the first cell, indicating different coupling with the incoming plane wave. 

If we modify the crystal lattice to align successive rows of scatterers along the propagation direction, thus forming a rectangular lattice, no dual-mode behavior is observed. Interestingly, the dual-mode behavior can be recovered if we consider a square lattice oriented along the ${\Gamma}M$ direction (i.e., along the square's diagonal). These cases show that the effect is strongly related to the geometry of the lattice, and more particularly to the alignment of the rods between successive rows of the crystal. Due to the quadrupolar resonance, the critical coupling directions are along and perpendicular to the propagation direction. Thus, in configurations where the rods in successive rows are staggered, two solutions can coexist at the same frequency, which mainly differ by the phase shift between successive rows, but are similar in terms of phase shift over two rows. This interpretation is reinforced by the proximity of the two modes to the $M$ and $\Gamma$ points in the multimode frequency band (successive rows out-of-phase and in-phase, respectively). For structures where rods in successive rows are aligned, the strong coupling between the resonant fields in adjacent rows only allows for one phase shift at a given frequency, giving rise to only one complex dispersion branch inside the gap.

In conclusion, using decomposition methods based on experimental time-dependent transmitted signals, as well as numerically obtained field maps, we have demonstrated that the unusual dispersion properties of this 2D fluid/solid crystal can be explained by the presence of two competing propagation modes. Our detailed analysis reveals the conditions under which such atypical dispersion characteristics occur:  simultaneous Bragg and hybridization mechanisms, a crystal structure in which the inclusion positions are staggered in adjacent rows perpendicular to the propagation direction (such as in a triangular lattice) and a resonance symmetry that allows competing interactions between inclusions on neighboring and next neighboring rows (such as quadrupolar resonances).  Thus, our results lay the foundation for future work with other phononic and even photonic crystals, opening opportunities for practical applications that exploit such two-mode behavior.  For example,  since the transmission spectra depend strongly on the balance between the two propagation modes, both in amplitude and phase, this type of crystal could have interesting applications for signal dispersion control. The full potential of this type of application depends on the ability to tune the resonant frequencies of the inclusions, a situation that is readily fulfilled for the model nylon rod-based case studied here, since the properties of nylon enable several crystal tuning mechanisms (temperature, water absorption and tension).  Some of these mechanisms could even be implemented rapidly enough (e.g., by embedding fine heating wire inside the rods or by modulating the tension) to enable dynamic processing control. 

Funding from NSERC is gratefully acknowledged.

\bibliography{PnC20bib}

\end{document}